\documentclass[12pt]{article}
\pdfoutput=1

\usepackage{graphicx}

\def\Msun{{\rm M}_\odot}
\def\Hmol{\ensuremath{\mathrm{H}_2}}
\def\gcc{\ensuremath{\mathrm{g}~\mathrm{cm}^{-3}}}
\def\pcc{\ensuremath{\mathrm{cm}^{-3}}}
\def\apj{\rm{ApJ}}
\def\mnras{\rm{MNRAS}}%
\def\apjl{\rm{ApJ}}%
\def\apjs{\rm{ApJS}}%

\usepackage{scicite}

\newenvironment{sciabstract}{%
\begin{quote} \bf}
{\end{quote}}

\newcounter{lastnote}
\newenvironment{scilastnote}{%
\setcounter{lastnote}{\value{enumiv}}%
\addtocounter{lastnote}{+1}%
\begin{list}%
{\arabic{lastnote}.}
{\setlength{\leftmargin}{.22in}}
{\setlength{\labelsep}{.5em}}}
{\end{list}}

\title{The Formation of Population III Binaries from Cosmological Initial Conditions}

\author
{Matthew J. Turk,$^{1\ast}$ Tom Abel,$^{1}$ Brian O'Shea$^{2}$\\
\\
\normalsize{$^{1}$Kavli Institute for Particle Astrophysics and Cosmology,
Stanford University, }\\
\normalsize{2575 Sand Hill Road, Menlo Park, CA 94025,}\\
\normalsize{$^{2}$Department of Physics \& Astronomy, Michigan State University, East Lansing, MI 48824-2320}\\
\\
\normalsize{$^\ast$To whom correspondence should be addressed; E-mail:
mturk@slac.stanford.edu.}
}

\date{}

\begin{document} 

\maketitle 

\begin{sciabstract}
Previous high resolution cosmological simulations predict the first stars to
appear in the early universe to be very massive and to form in isolation.  Here we
discuss a cosmological simulation in which the central $50\Msun$
clump breaks up into two cores, having a mass ratio of two to one, with one fragment
collapsing to densities of $10^{-8}\gcc$.  The second fragment, at a distance of
$\sim800$ astronomical units, is also optically thick to its own cooling radiation from
molecular hydrogen lines, but is still able to cool via collision-induced emission.
The two dense peaks will continue to accrete from the surrounding cold gas
reservoir over a period of $\sim10^5$ years and will likely form a binary star system.
\end{sciabstract}

Hydrodynamical simulations that start from cosmological initial conditions have
predicted that the first luminous objects in the universe were isolated stars with
masses in the range $30-300\Msun$, based on accretion rates calculated in the absence
of accretion-inhibiting factors \cite{ABN02,2008Sci...321..669Y,2004NewA....9..353B}.
Idealized protostellar evolution simulations suggest accretion should end when the star
reaches $\sim100\Msun$ \cite{2003ApJ...589..677O}, but even though most relevant
feedback mechanisms of the protostars on their accretion flow are known
\cite{2008ApJ...681..771M}, no fully self-consistent radiation hydrodynamical
simulations reaching the main sequence have yet been possible, even in one dimension.
The early stages of proto-stellar evolution, however, are understood in some detail
from spherically symmetric calculations \cite{1998ApJ...508..141O,2004MNRAS.348.1019R}
as well as semi-analytic models \cite{TanMcKee2004}.

The environment in which the first stars form is calculated by following
early-universe evolution of the primordial gas and dark matter; by this means, the
first stars are found to form in halos with a total mass of $\sim10^6\Msun$, with
collapse driven first by cooling via molecular hydrogen ro-vibrational lines and later
by collision induced emission \cite{1998ApJ...508..141O,RA04,2008AIPC..990...25G}.
While several tens of calculations \cite{ABN02,oshea07a,2006ApJ...652....6Y} have
followed the collapse of a primordial protostellar halo to relatively low densities
($\sim10^{-12}\gcc$), only two previous calculations have followed the collapse from
cosmological initial conditions to protostellar densities
\cite{2008Sci...321..669Y,2008AIPC..990...16T}.  Both of these calculations utilized a
chemical model that included the relevant medium- and high-density physics, such as
heating from the formation of molecular hydrogen, collisionally induced emission,
three-body molecular hydrogen formation, and gas opacity at high densities.
Several simulations have followed the collapse of a single metal-free cloud starting
from idealized initial conditions
\cite{2002ApJ...564...23B,2004NewA....9..353B,2001ApJ...548...19N}.  Parameter studies
of rotating cylinders have shown that metal-free gas can fragment
\cite{2004ApJ...615L..65S} and have found fragmentation of spherical distributions of
extremely low- and zero-metallicity gas \cite{2008ApJ...672..757C}, including in 3D
nested grid parameter studies \cite{2008ApJ...677..813M}.  Additionally, studies of
zero-metallicity gas collapsing in isolation have found fragmentation at low densities
($\sim10^{-16}~\gcc$) \cite{1999ApJ...527L...5B}.  However, no previous simulation
starting from cosmological initial conditions has shown fragmentation in a cosmological
context.  

We present simulations performed with the adaptive mesh refinement code Enzo
\cite{2004astro.ph..3044O}.  These simulations were initialized at a redshift of $99$ in a box
$300~\mathrm{kpc}~\mathrm{h}^{-1}$ (comoving).  (Detailed simulation parameters and
methods can be found in the supporting online material.)  We stopped the simulation at
a maximum (proper) baryon density of $1.61\times 10^{-8}\gcc$ at $z=19.08$ (189 million
years after the Big Bang) and $1.72~R_{\odot}$ (proper) peak spatial resolution, where we
have resolved the first massive halo that forms in this simulation volume.  The
simulation presented here is explicitly a simulation of the first generation of
Population III star formation.  The second generation of Population III stars are still
primordial in composition but are affected by previous generation of stars, potentially by
effects such as kinetic energy injection, radiation backgrounds or cosmic rays.  We
have simulated five realizations of first-generation Population III stars that evolve
to at least this density and have found fragmentation in one.  The dark matter halo
merger history and the evolution of baryonic properties at redshifts above 20 are
typical of previous simulations \cite{ABN02,2003ApJ...592..645Y}.

The collapsing halo fragmented at a density of $\simeq2.0\times10^{-13}\gcc$ at
redshift $z=19.08$, roughly. We have conducted resolution studies, where we found
fragmentation but a change in the separation of the identified cores.  (More
information can be found in the supporting online materials.)  To study the
fragmentation of the cloud, we identified gravitationally-bound,
topologically-connected regions at or above a single density within the halo's virial
radius. The density at which these regions diverged was $\sim2.0\times10^{-13}\gcc$,
however, because the difference was only a thin strip of lower density gas, we
conducted all subsequent analysis on connected isodensity surfaces with minimum density of
$3.0\times10^{-13}\gcc$.  We selected material within two isodensity surfaces,
hereafter referred to as Core~A and Core~B ({\it Fig.~1}).  Core A is the more massive core,
with a mass of $10.0\Msun$, and Core~B of $6.3\Msun$.  Determining whether these two
objects will ultimately merge is nontrivial, but estimates can be made based on the
current orbital mechanics as well as collapse conditions of each core in isolation.  If
these two objects form hydrostatically-supported protostellar cores and most of the
intervening gas is accreted onto them, they will remain separate and form a binary star
system.

Both cores are oblate and are embedded in a crescent-shaped cloud.  Even though they
can not be reliably identified more than $\sim175~\mathrm{years}$ before the end of the
simulation, the beginning of fragmentation is already evident in both density and
molecular hydrogen, where lower-density, mostly-atomic regions are permeated by
filaments of higher-molecular fraction gas.  ({\it Fig.~1}, top row.) Averaging over the
time period during which the two cores are identifiable and distinct, Core~A
experiences $0.061~\Msun / \mathrm{year}$ mass flux across its density isosurface,
while Core B has mass flux of $0.049~\Msun / \mathrm{year}$.  ({\it Fig.~2})  The flow
across both density isosurfaces is roughly linear.  The peak densities in both cores
evolve at close to free-fall trajectories and Core~B collapses $\sim60$ years after
Core~A.

The separation of the centers of mass of the two cores is 800AU.  Within a minimally
enclosing sphere of radius 780AU located at the center of mass of the two cores, we
find a total mass of $52\Msun$; within a sphere of radius twice the separation of the
cores, we find a mass of $99~\Msun$.  By defining $$ \bar{R} = \left({\frac{3
M_{\mathrm{enc}}}{4\pi \bar{\rho}}}\right)^{1/3} $$ we calculated the characteristic
radius of each core.  Core~B, the lower-mass core, is at this time less dense then
Core~A, and has a smaller characteristic radius.  Given its current angular
momentum and assuming both angular momentum conservation and no material between the
cores its final Keplerian radius is given by $$ R_{\mathrm{kep}} =
\frac{L_B^2}{GM_{\mathrm{A}}} $$ where $L_B$ is the specific angular momentum of Core B
in the rest frame of Core A and $M_\mathrm{A}$ is the mass of Core A.
$R_{\mathrm{kep}}$ evaluates to $2400~\mathrm{AU}$, which is three times their current
distance, indicating that the velocity of Core B is super-Keplerian.  The tangential
velocity of Core~B with respect to Core~A is $5.9~\mathrm{km}~\mathrm{s}^{-1}$, and it
has a radial velocity of $-5.3~\mathrm{km}~\mathrm{s}^{-1}$.  Given the current
velocities, separation and masses of the cores and the assumption that the two cores
exist with their current trajectories in a vacuum, a simple orbit integration shows
that their closest approach of 480 AU would occur after 360 years, more than 10
dynamical times of the lighter core.  However, this vacuum assumption could be
misleading. Another limit can be derived assuming the remaining $46\Msun$ mass
between Core~A and B instantaneously were accreted onto only Core~A, then the orbit
integrations show it would take over 400 years for B to come within 300 AU of A.
Because the time scale of the formation of hydrostatically-supported, protostellar
cores in either core is close to the free fall time, an accreting binary system will
form.  Following the formulation of the dynamical friction force on a massive perturber
in a near-circular orbit \cite{1999ApJ...513..252O}, the loss of angular momentum
through dynamical friction does not substantially increase the likelihood of a merger
of the two cores; in fact, by providing an upper bound on the infall rate, we see that
both cores will form protostars long before merging.  We have calculated the average
density of the background medium using spheres both two and five times the radius of
the core separation and found that the two cores should not approach closer than
$100~\mathrm{AU}$ for at least $3600~\mathrm{years}$.

We plotted radially-binned, mass-weighted average quantities at several epochs of the
halo collapse.  (Fig.~3)  The central bin was located at the most dense zone in the
simulation at all times.  After the cloud fragmented, this most dense zone was always
within Core~A.  The location of Core~B is clearly visible as a spike in the density,
temperature, and \Hmol~fraction, and a slight enhancement in the enclosed mass, at
$\sim800\mathrm{AU}$.  The temperature plot shows a marked increase at high densites
compared to that of both previous studies \cite{ABN02,oshea07a}.  This is likely a
result not only of different molecular hydrogen formation and cooling rates, and also
the inclusion of the heating from the formation of molecular hydrogen and opacity to
molecular hydrogen lines.  These factors delay the conversion to a fully-molecular
state until the gas collapses to higher densities, reducing the efficiency of the
ro-vibrational cooling, leading to higher temperatures.  Furthermore, the average
molecular hydrogen fraction of the cloud does not reach unity, as it has in previous
calculations.  At higher densities the cooling from ro-vibrational lines is inefficient
as a result of increased optical depth; this leads to an increase in the temperature,
and thus a drop in the molecular hydrogen fraction.  There is no substantial
dissociation at the final output of this simulation, but the inner region of the
collapsing halo has a molecular hydrogen mass fraction of $\sim0.5$.  As the cloud
collapses further, \Hmol~will dissociate at densities just slightly higher than those
presented here.  Following the prescription of \cite{RA04}, the peak density of the
simulation provides an effective transmission coefficient of $2\times10^{-3}$ for
ro-vibrational cooling and a transmission coefficient of $0.94$ for cooling from
collision induced emission.

To estimate instabilities brought on by chemical and radiative mechanisms, we follow
the methods of \cite{RA04}; when the characteristic timescales for the change in energy
of the gas is less than the dynamical time, small perturbations can no longer be
effectively suppressed and the gas is susceptible to fragmentation.  The dynamical time
is defined by $$ t_{\mathrm{dyn}} \equiv \sqrt{\frac{3\pi}{16 G \rho}} $$ By defining
$E$ as the total thermal energy of the gas, $\dot{e}_{\mathrm{rad}}$ as the radiative
cooling from ro-vibrational lines and collision-induced emission, and
$\dot{e}_{\mathrm{chem}}$ as the heating or cooling due to chemical changes in the
molecular hydrogen content of the gas ($4.48~\mathrm{eV}$ per molecule), we can write
the molecular hydrogen timescale as $$t_{\Hmol} = \frac{E}{\dot{e}_{\mathrm{rad}} -
\dot{e}_{\mathrm{chem}}}$$ Comparing these two quantities (Fig.~4) we found that the
ratio was approximately unity at most densities, but that at the densities at which the
gas fragmented, this ratio changed sign abruptly.  This is a steep transition, from being
moderately prone to fragmentation, to being dominated by the heating from the formation
of \Hmol.  However, as shown by \cite{2003ApJ...599..746O}, in isolation the thermal
instability brought on by the onset of three-body heating and molecular line cooling
would not be sufficient to cause the fragmentation of a collapsing gas cloud.

We define the rotational energy as $$ E_{\mathrm{rot}} \equiv \frac{1}{2}
\mathbf{\omega}\mathbf{I}\mathbf{\omega} $$ where $\mathbf{\omega}$ is the rotational
velocity and $\mathbf{I}$ is the inertial tensor of the gas.  For a sphere of radius
$r$ and mass $M$ the gravitational binding energy $W$ is $$ W \equiv
\frac{3}{5}\frac{GM^2}{r} $$ We compared these two quantities (Fig.~4, bottom) and
indicated with thin lines the critical ratios of $\sim0.44$ \cite{1988ApJS...66..315H}
and $0.27$ \cite{1987ApJ...323..592H} above which dynamical instabilites can become
dominant in compressible toroids and compressible spheroids with a
centrally-concentrated mass.  This ratio was at a maximum at an enclosed mass of
approximately $100\Msun$.  The transition between instabilities identified above is
within this radius.  This halo fragments into multiple components because of the
nesting of this marginal chemical instability within the radius of that of the
rotational instability.  However, more important is if an increase in this ratio is
found exactly at the radii where the cloud has been identified as fragmenting.  While
analytic estimates for the gravitational potential energy of a cloud vary by factors of
order unity depending on the eccentricity and ellipticity of the cloud, the relative
ratio traces the mass scales where fragmentation may occur.

The total number of first-generation Population III stars is likely to be significantly
smaller than that of the second generation.  Assuming no recombination, massive
metal-free stars can ionize of order $10^7~\Msun$ of gas, which is between $100-1000$ times
the gas mass of a Population III star-forming halo\cite{2007ApJ...659L..87A}, and thus will
be outnumbered by a large factor by metal-free stars forming in preprocessed regions.
Fragmentation in unprocessed, first-generation Population III stars thus indicates that
fragmentation in preprocessed, second-generation Population III stars is also possible.
This would ultimately be more important to structure formation and the star formation
history of the universe and the chemical signature of metal free stellar evolution in
the fossil record of old stars in the Galaxy.\cite{2006ApJ...641....1T}
By changing the size of the gas reservoirs from which the first-generation Population
III protostars stars accrete, as well as the ultraviolet flux from the subsequent
metal-free stars themselves, the long-term star formation environment for the next
generation of Population III stars could be dramatically changed.  A higher flux of
ionizing photons could lead to more efficient cooling by molecular hydrogen in
neighboring halos \cite{2005ApJ...628L...5O,2008ApJ...685...40W}
but a lower meta-galactic flux would likely lead to a higher global star formation rate
and higher accretion rates. \cite{2008ApJ...673...14O}

We have shown that in a particular realization fragmentation occurs at relatively high
densities.  The separation and timescales are such that the two identified fragments
are likely to form a binary stellar system.  Improvements to chemical solvers at high
densities and better laboratory values for the molecular hydrogen rate coefficients may
change the details of the collapse, but these results demonstrate that fragmentation is
possible in metal-free halos collapsing from realistic initial conditions.  The
frequency of such fragmentation, and thus that of metal-free binary systems, cannot be
gauged using the small number of simulations conducted thus far.  

The problem of ``finding'' fragmentation in cosmological halos may be one of poor
sampling; if fragmentation is rare, the small number of published calculations likely
will not sample those halos in which it could occur.  In particular, if fragmentation
is more likely to occur in halos that undergo rapid merger history, the current means
of ab initio simulation of Population III star formation may be ill-equipped to study
its likelihood and relevance.  We have found fragmentation in one out of five
realizations, suggesting that binaries are a formation channel that must be considered
in population synthesis studies.  From a set of only five realizations an initial mass
function cannot be estimated.  However, we conclude by noting that the current means of
conducting high-dynamic range simulations are biased against finding
fragmentation in collapsing pre-stellar clouds.  This simulation fragmented at a
density of $\simeq2.0\times10^{-13}\gcc$, which corresponds to a dynamical time of 210
years; if the time delay between clump formation is greater than approximately the
dynamical time, the current methods of simulating primordial star formation will not
observe fragmentation.  Consequently it is likely that a substantial fraction of
Population III stars are forming binaries or event multiple systems.

\clearpage
\includegraphics[width=0.88\textwidth]{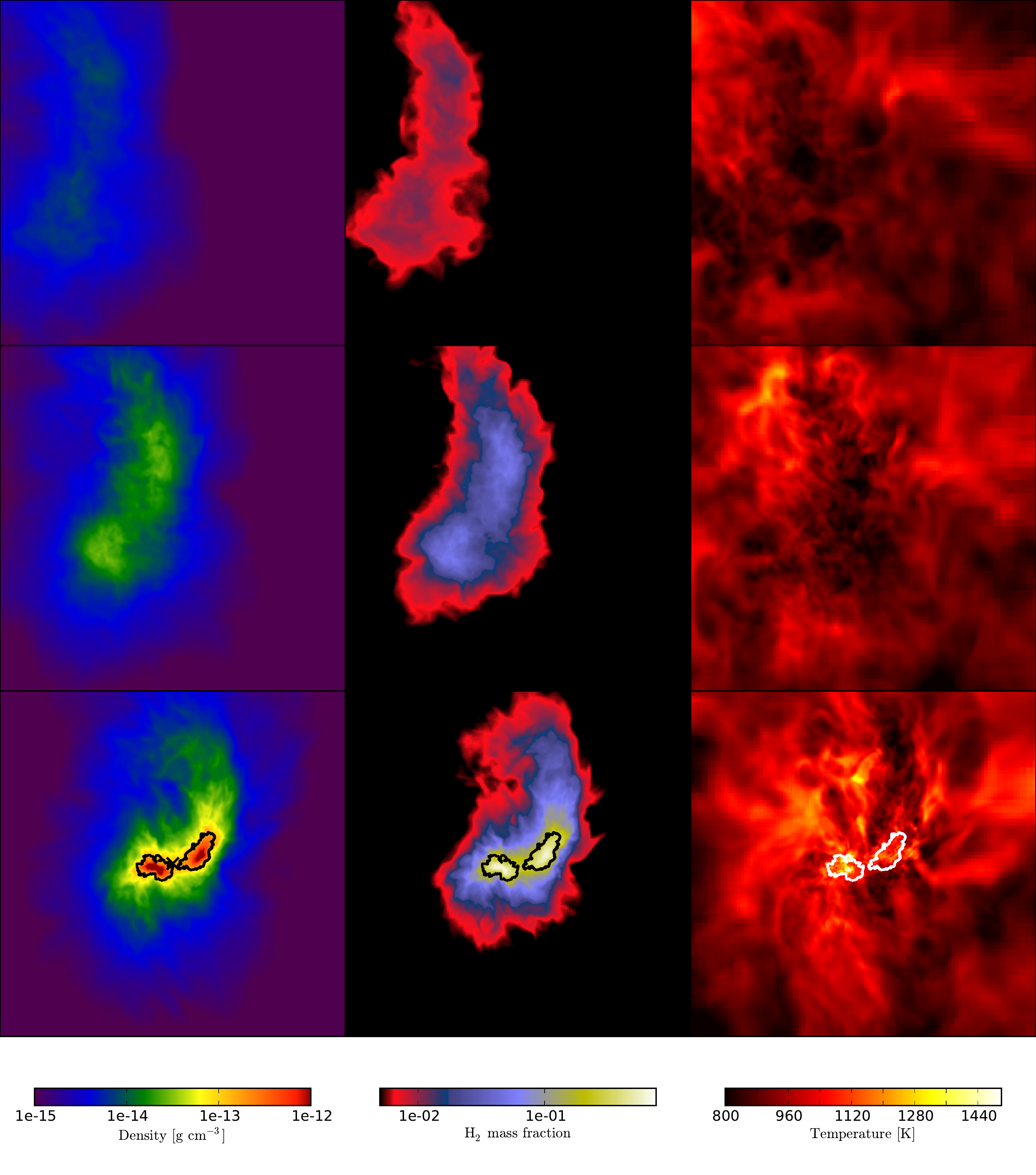}
\noindent{\bf Fig. 1.} Mass-weighted average density (left column), $\mathrm{H}_2$ fraction
(middle column), and temperature (right column) projected through a cube centered on
the center of mass of the two-core system with a side length of $3500~\mathrm{AU}$.
The bottom row is the final output of the simulation, the middle row is 555 years
previous, and the top row is an additional 591 years previous to the middle row (1146
years before the end of the simulation).  Gravitationally bound cores with minimum
density $3.0\times10^{-12}\gcc$ have been outlined with thick lines in the bottom row;
Core~A is on the left and Core~B is on the right.  Field of view is 3500AU.

\clearpage
\includegraphics[width=0.88\textwidth]{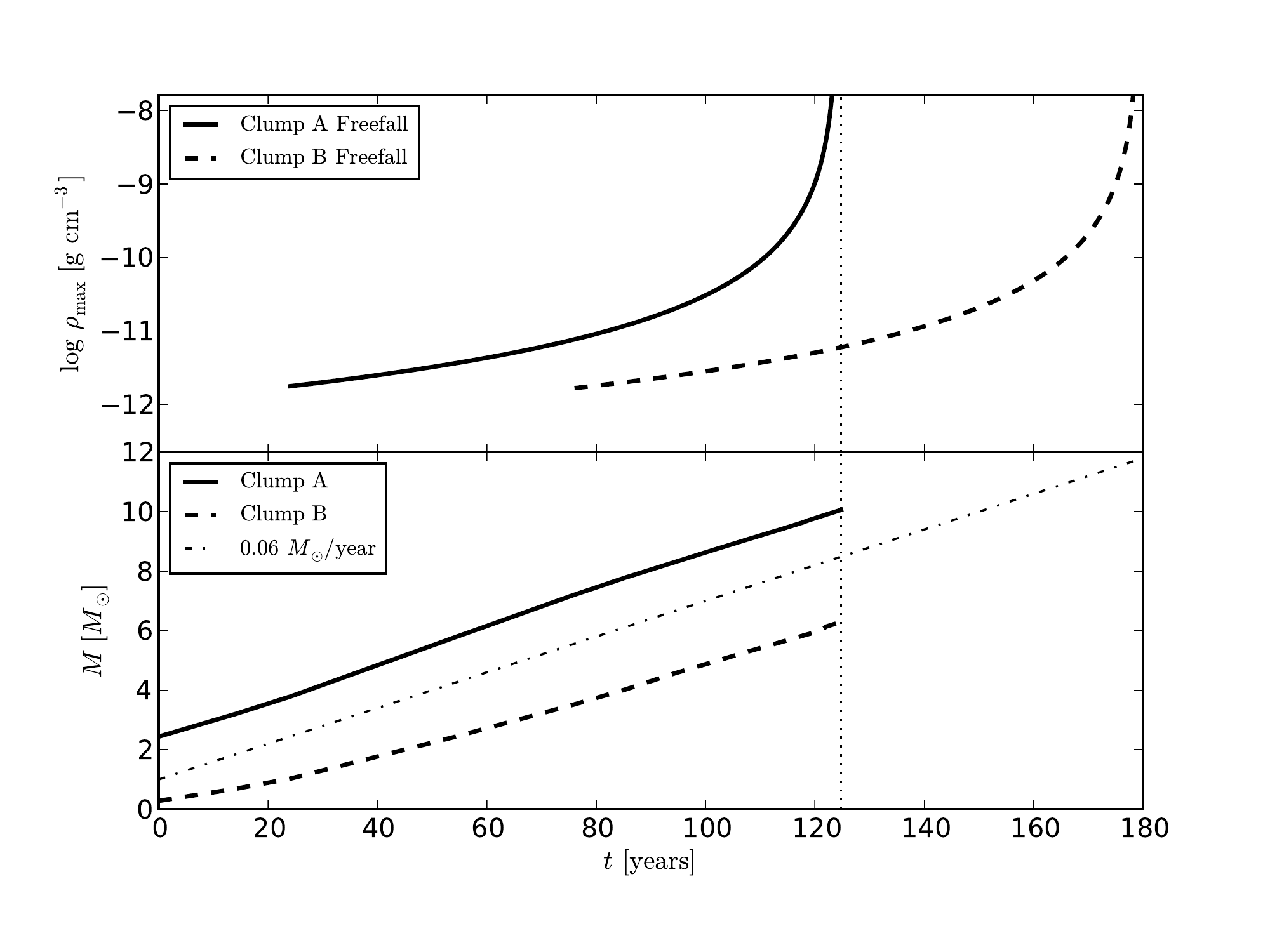}
\noindent{\bf Fig. 2.} Cores, identified by placing a minimum density cut at
$3.0\times10^{-13}\gcc$, plotted over time.  The y-position denotes current maximum
density, x-position denotes time since the first core was identified.  Overlaid is the
predicted free-fall evolution of the peak density in each core starting from the first
time when the maximum density is greater than $10^{-12}\gcc$, governed by
$\frac{d\rho}{dt} = \frac{\rho}{t_{ff}}$, where we use $ t_{ff} =
\sqrt{\frac{3\pi}{32G\rho_{\mathrm{max}}}} $.  We have extended the free-fall
trajectories past the end of the simulation, which is indicated by the dotted line.

\clearpage
\includegraphics[width=0.88\textwidth]{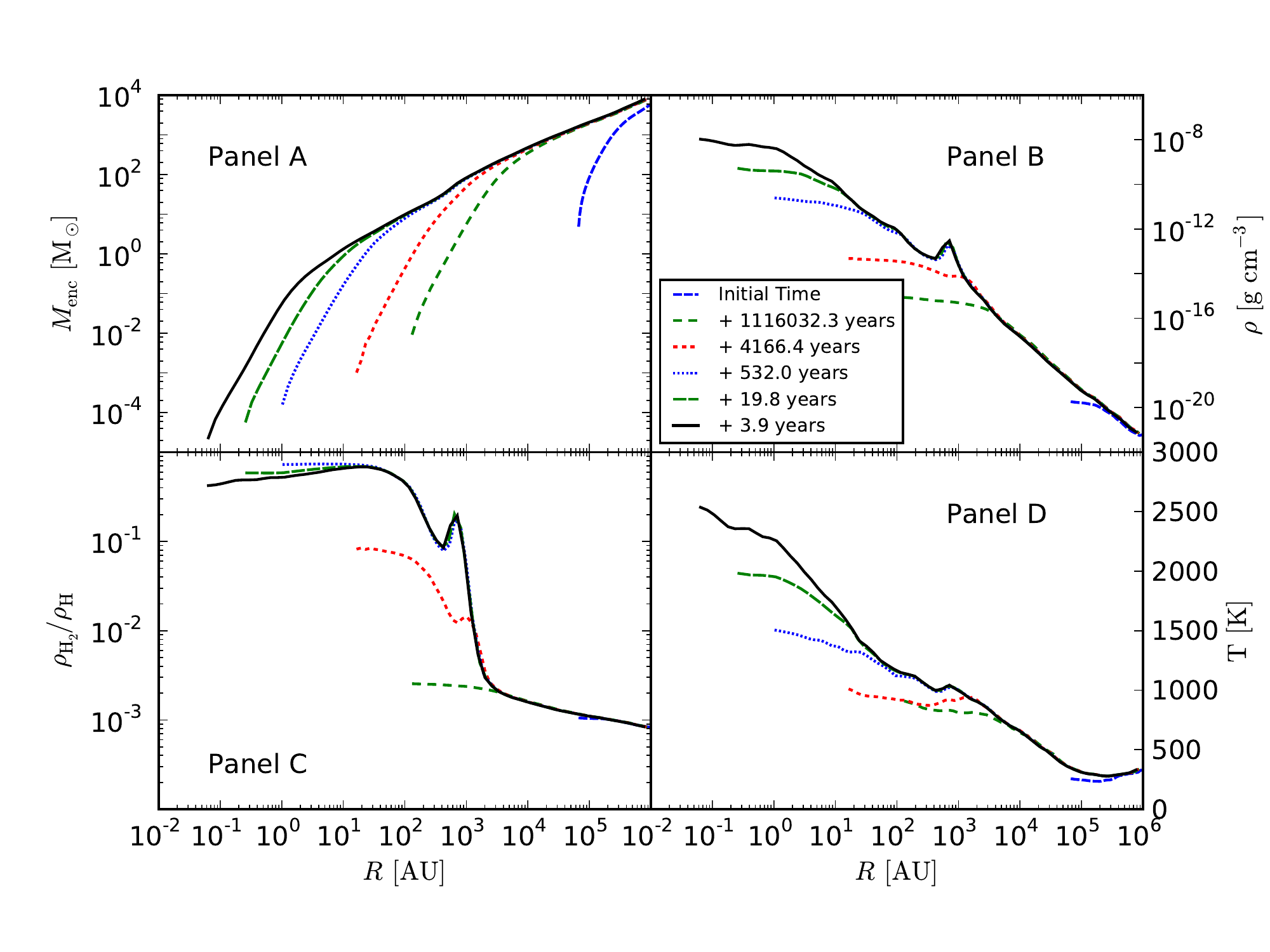}
\noindent{\bf Fig. 3.} Mass-weighted, spherically-averaged quantities as a function of distance from
the most dense point (which is located within Core~A after the cloud fragments) at
different times in the simulation: mass-enclosed as a function of radius (Panel A),
density as a function of radius (Panel B), molecular hydrogen mass fraction (Panel C),
and temperature (Panel D).

\clearpage
\includegraphics[width=0.88\textwidth]{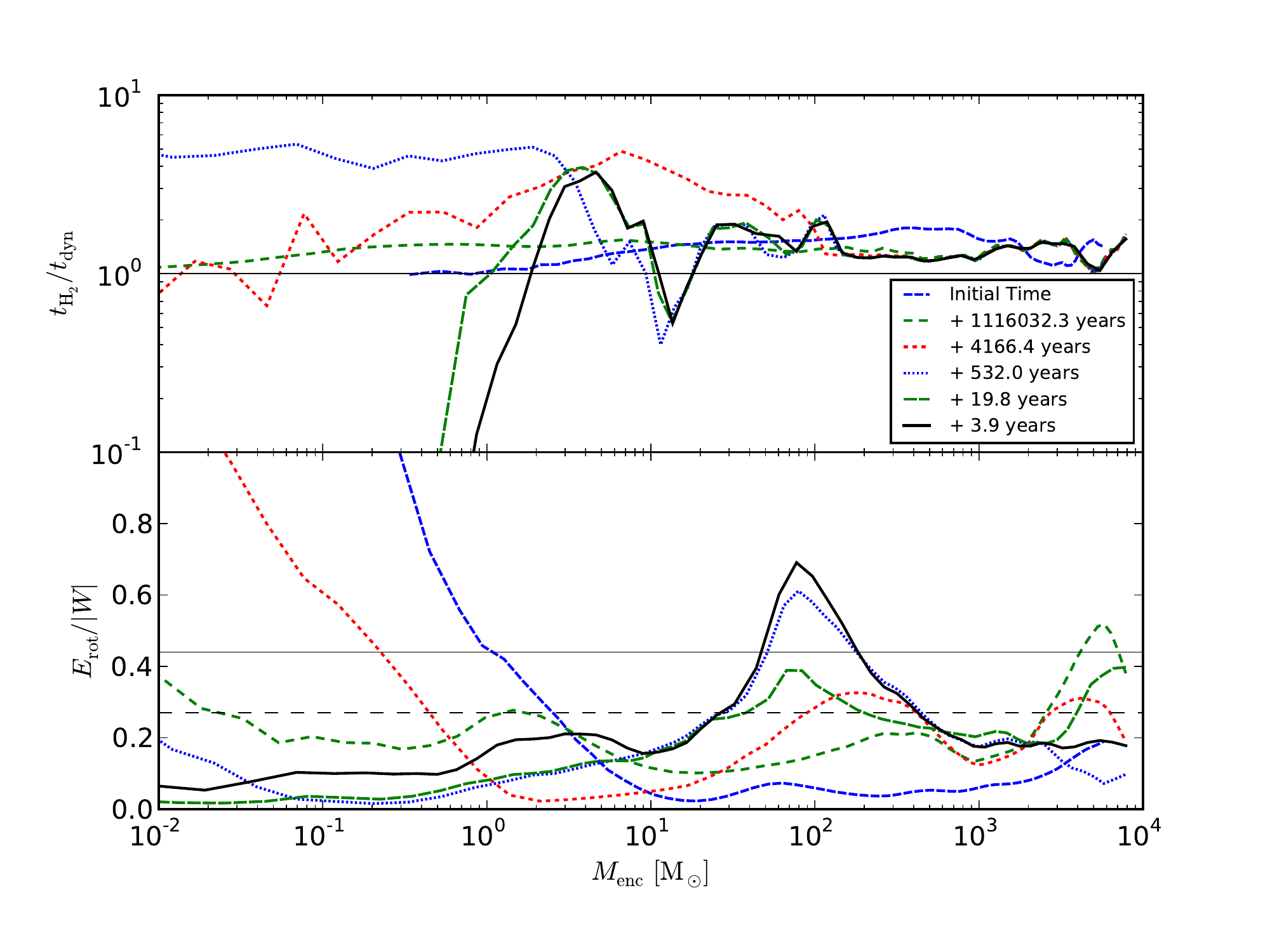}
\noindent{\bf Fig. 4.} Enclosed quantities as a function of mass-enclosed, with respect
to the most dense point (which is located within Core~A after the cloud fragments) and
calculated in the rest frame of that point at different times in the simulation: the
mass-weighted average ratio of the dynamical time of the gas divided by the
cooling time of the molecular hydrogen, taking into account the heating
from three-body formation processes (top panel), and rotational energy divided by
gravitational binding energy (bottom panel).  In the bottom panel, lines have been
drawn to indicate a ratio of 0.27 (thin, dashed) and 0.44 (thin, solid).

\section*{Supporting Online Material}

\subsection*{Simulation Methodology}

The simulation presented in this work corresponds to simulation
L0\_30C in \cite{oshea07a}, which was initialized from Gaussian random initial
conditions at redshift $z=99$ inside a cubic volume with a side length of
$300~\mathrm{kpc}~\mathrm{h}^{-1}$ (comoving).  In addition to standard refinement
criteria based on overdensity and cooling time, we require a minimum of 64 cells
per Jeans length, ensuring adequate resolution at all stages of collapse,
ending with 31 levels of refinement.  

The simulations presented here implement a primordial chemistry solver
optimized to be efficient in the stiff regime when three-body reactions forming
molecular hydrogen become dominant, utilizing the method from
\cite{Verwer94gauss-seideliteration}.  The equation of state of molecular
hydrogen is also treated accurately, and all kinetic rate equations are solved
without applying prescriptions for equilibria.  In contrast to previous
adaptive mesh refinement simulations, these calculations include energy
contributions due to the formation and destruction of molecular hydrogen as a
result of three-body reactions ($4.48~\mathrm{eV}/\mathrm{molecule}$), the
dominant \Hmol~formation channel above densities of $10^{-16}\gcc$.  These
energy contributions, as well as energy losses due to radiative processes, are
split into formation and destruction contributions and the energy is updated in
a second-order backward difference method.  This allows for fast numerical
convergence during integration.  The changes in the gas chemical state are
updated on the same timescale as the thermal changes, ensuring accurate
chemothermal composition.

For the coefficients governing the formation and dissociation of molecular
hydrogen,

$$\begin{array}{llcl}
(k_{22}) & \mathrm{H} + 2\mathrm{H} & \rightarrow & \mathrm{H} + \mathrm{H}_2 \\
(k_{13}) & \mathrm{H} + \mathrm{H}_2 & \rightarrow & \mathrm{H} + 2\mathrm{H}
\end{array}$$

we use the rate calculated in \cite{2008AIPC..990...25G} where a discussion of
the uncertainty governing this particular process is detailed.  This
uncertainty is further discussed in \cite{2008MNRAS.388.1627G}.  We have also
used a prescription from \cite{RA04} for the optical depth of ro-vibrational
cooling from molecular hydrogen, as well as the optical depth to
collision-induced absorption.  We have additionally included the formation of
molecular hydrogen via three-body reactions with \Hmol~as the third body, using
the rates given in \cite{PSS83}.  The remaining set of chemical rate
coefficients is identical to that of \cite{oshea07a}.

We have simulated five different cosmological halos, with four of the
calculations corresponding to to runs from \cite{oshea07a}.  The simulation
presented in this work corresponds to simulation L0\_30C in \cite{oshea07a},
which was initialized from Gaussian random initial conditions at redshift
$z=99$ inside a cubic volume with a side length of
$300~\mathrm{kpc}~\mathrm{h}^{-1}$ (comoving).  The other four realizations did
not fragment before we stopped each simulation.  We initialize the simulation on a
$128^3$ root grid with three nested levels of refinement, for an effective resolution
of $1024^3$ ($0.42$ kpc proper at $z=99$).  Our finest dark matter particle mass is
$2.6\Msun$.  The first protostar forms at redshift $z=19.08$, inside a dark matter halo
of mass $5.8\times10^{5}~\mathrm{M}_\odot$ with a gas spin parameter of $0.025$ and
dark matter spin parameter of $0.042$, where the spin parameter is defined as:
$$
\lambda = \frac{L|E|^{1/2}}{GM^{3/2}}
$$
where $L$ is the mass-weighted average specific angular momentum, $E$ is the
total kinetic energy and $M$ is the enclosed mass.  This
collapse time is somewhat delayed from that in \cite{oshea07a}, which we
attribute to different choices of molecular hydrogen formation rate and cooling
rates; however, the two simulations are in general agreement.  The simulation
was stopped at a maximum (proper) baryon density of $1.61\times 10^{-8}\gcc$
($n_{\mathrm{H}}\sim10^{16}\pcc$), at a redshift of $19.08$, with 31 levels of
refinement.  The peak spatial resolution is $1.72~R_{\odot}$ (proper) and at
the end of the simulation there were $8.75\times10^{7}\approx 440^3$ unique
computational elements with a final mass resolution of $\approx
10^{-7}~M_{\odot}$.

\subsection*{Quantities}
We tabulate several relevant quantities of the two cores in
Table~\ref{clump_table}.  These values have all been calculated during the
final output from the simulation, when the maximum baryon density was
$1.61\times10^{-8}\gcc$.  

\begin{table}[p]
\begin{centering}
\begin{tabular}{lrrrrr}
Core & $\rho_{\mathrm{max}}$ & Mass & $\bar{R}$ & $t_{ff} $ \\\hline
A & $1.61\times10^{-8}\gcc$ & $10.0\Msun$ & 24.4 AU & 0.52 years \\
B & $5.03\times10^{-12}\gcc$ & $6.3\Msun$ & 98 AU & 29.7 years
\end{tabular}
\caption{Table of quantities for identified gas cores from final simulation
output}
\label{clump_table}
\end{centering}
\end{table}

\subsection*{Resolution Studies}

We have conducted resolution studies of this particular realization.
With the refinement criteria such that we had 16 cells per Jeans
length (J16), we allowed the simulation to run until the maximum density was
$10^{-5}\gcc$, where we observed that two cores had formed with a separation
of $\sim200~\mathrm{AU}$ and masses of two and seven solar masses, with
topologically distinct, gravitationally-bound density contours at a density of
$9.5\times10^{-13}\gcc$.  The separation was found to be lower than the
Keplerian radius.  To ensure that this fragmentation was not the result of
resolution-limited fragmentation, we restarted the simulation from the time
when the maximum density was $10^{-21}\gcc$, before the onset of rapid
molecular hydrogen formation.  With a total of 32 cells per Jeans length (J32), we
saw fragmentation at length scales of $\sim100~\mathrm{AU}$.  However, because
of the close pair distance, it is unclear whether or not the cores would
subsequently merge.  The work presented here had the greatest resolution,
requiring 64 cells per Jeans length (J64).

By changing the required resolution, we have changed the structure of the
hydrodynamic flow into the collapsing region.  The delay time in the simulation
corresponded roughly to the strength of fragmentation; J32 collapsed the
earliest, with J16 collapsing 6000 years later and J64 collapsing 175,000 years
following that.  This corresponds to roughly $8\%$ of the freefall time at the
density at which we restarted the simulation.  The actual collapse takes about
5 million years because of the balance of molecular cooling and heat input from
turbulence on the large scales ($\sim\mathrm{pc}$) as hitherto seen and
explained by \cite{2003ApJ...592..645Y}.  As we conduct the resolution study
varying the spatial resolution by a factor 4 one also slightly modifies the
``initial conditions'' since we start the calculation not shortly after
recombination, where the original simulation was initialized, but rather at an
intermediate time.  The loitering phase relevant for primordial star formation
otherwise will always be prone to the intermittency issues known so well in
interstellar turbulence.  Nevertheless, looking at the difference between the
J16 and J32 runs the slight change in ``initial conditions'' only lead to a
6000 year difference.  The timing clearly is the most significant differences
in these runs.

\subsection*{Core Finding Algorithm}

We identified cores by using an optimized version of the core finder
described in \cite{2009ApJ...691..441S}.  Starting with a sphere of radius
$2500 \mathrm{AU}$ in radius, centered at the most dense point in the
simulation, we recursively identify topologically connected sets of cells of a
given density minimum, incrementing the density minimum with every subset of
the data that is extracted.  All sets of cells that are not gravitationally
bound are eliminated, as are objects with only a single subset identified.
This allows us to identify the largest topologically connected sets of cells
that are gravitationally bound.

In this algorithm we compute the gravitational energy by direct summation,
rather than using the potential Enzo computes.  Consequently the computational
cost scales with $N^2$ where $N$ is the number of cells in the simulation.  As
such, performing this computation on commodity CPUs became cost prohibitive; to
that end, we converted the code to run on the NVidia CUDA
framework\footnote{\texttt{http://www.nvidia.com/cuda/}}.  This decreased
computational time substantially, allowing us to perform finely-spaced sweeps
of density-space (separated by $0.15$ dex), ensuring we have accurately
identified fragments.

\begin{scilastnote}
\item We thank V. Bromm, G. Bryan, A. Escala, S. Glover, C. McKee, J. Oishi, 
B. Smith, J. Tumlinson, and N. Yoshida for useful discussions. This  
work was partially supported by the U.S. Department of Energy contract 
to SLAC no. DE-AC02-76SF00515, NASA ATFP grant NNX08AH26G and NSF  
AST-0807312. BWO and MJT carried out this work in part under the  
auspices of the National Nuclear Security Administration of the U.S.  
Department of Energy at Los Alamos National Laboratory under Contract 
No. DE-AC52-06NA25396. BWO was partially supported by a LANL  
Director*s Postdoctoral Fellowship (DOE LDRD grant 20051325PRD4).
\end{scilastnote}

\end{document}